\documentclass[aps,prl,twocolumn,groupedaddress,showpacs]{revtex4}

\usepackage{graphicx}

\begin{document}

\newcommand{\etal}{\emph{et al.}}
\newcommand{\Li}{{}^{6}\textrm{Li}}
\newcommand{\K}{{}^{40}\textrm{K}}
\newcommand{\Rb}{{}^{87}\textrm{Rb}}
\newcommand{\kB}{k_{\mathrm{B}}}
\newcommand{\Rbmp}{\textrm{Rb}\left|2,2\right\rangle}
\newcommand{\RbTwoOne}{\textrm{Rb}\left|2,1\right\rangle}
\newcommand{\RbTwoZero}{\textrm{Rb}\left|2,0\right\rangle}
\newcommand{\RbOneZero}{\textrm{Rb}\left|1,0\right\rangle}
\newcommand{\RbOneOne}{\textrm{Rb}\left|1,1\right\rangle}
\newcommand{\RbOneMOne}{\textrm{Rb}\left|1,-1\right\rangle}
\newcommand{\Kmp}{\textrm{K}\left|9/2,9/2\right\rangle}
\newcommand{\Limp}{\textrm{Li}\left|3/2,3/2\right\rangle}
\newcommand{\KNinemNine}{\textrm{K}\left|9/2,-9/2\right\rangle}
\newcommand{\KNinemFive}{\textrm{K}\left|9/2,-5/2\right\rangle}
\newcommand{\LiOneOne}{\textrm{Li}\left|1/2,1/2\right\rangle}

\title{Ultracold Heteronuclear Fermi-Fermi Molecules}

\author{A.-C.~Voigt}\email[Electronic address: ]{arne.voigt@mpq.mpg.de}
\author{M.~Taglieber}
\author{L.~Costa}
\author{T.~Aoki}
\author{W.~Wieser}
\author{T.W.~H\"ansch}
\author{K.~Dieckmann}
\affiliation{Department f\"ur Physik der Ludwig-Maximilians-Universit\"at, Schellingstra\ss e\,4, 80799 Munich, Germany\\
Max-Planck-Institut f\"ur Quantenoptik, Hans-Kopfermann-Stra\ss e 1, 85748 Garching, Germany}

\date{\today}

\begin{abstract}
We report on the first creation of ultracold bosonic heteronuclear molecules of two fermionic species, $\Li$ and $\K$, by a magnetic field sweep across an interspecies s-wave Feshbach resonance. This allows us to associate up to $4\times 10^4$ molecules with high efficiencies of up to 50\%. Using direct imaging of the molecules, we measure increased lifetimes of the molecules close to resonance of more than $100\,\mathrm{ms}$ in the molecule-atom mixture stored in a harmonic trap.
\end{abstract}


\maketitle
Two-component mixtures of fermionic quantum gases have attracted much interest over the past years. In these systems, long-lived, weakly bound molecules were produced made up of two atoms of the same species in different internal states \cite{HomonuclearFermiMolecules}. The long lifetimes observed for these molecular gases are a consequence of the Pauli principle, which suppresses three-body collisions, and hence vibrational quenching, in a system of not more than two distinguishable components \cite{Petrov2004}. More recently, the interest has shifted towards ultracold \emph{hetero}nuclear diatomic molecules, which can have a large electric dipole moment \cite{Aymar2005}. So far, Bose-Bose \cite{HeteronuclearBoseBoseMolecules} and Bose-Fermi \cite{HeteronuclearFermiBoseMolecules} dimers have been produced. However, among the ultracold heteronuclear dimers the \emph{Fermi-Fermi} molecules are of special interest since they are expected to exhibit long lifetimes for the same reasons as in the homonuclear case \cite{Petrov2005}. Long-lived polar molecules open the door to the creation of a molecular Bose-Einstein condensate (BEC) \cite{MolecularBEC} with anisotropic, electric dipolar interaction and show potential for precision measurements \cite{PrecisionMeasurements} and novel quantum information experiments \cite{DeMille2002}. Furthermore, the two-species Fermi-Fermi mixture may allow the realization of novel quantum phases \cite{ManyBody, Petrov2007, QCD} and offers the possibility to tune interactions and to conveniently apply component-selective experimental methods.

In this Letter, we present the first production of ultracold diatomic molecules composed of two different fermionic atomic species. We study the creation process of the molecules, their lifetime in a molecule-atom mixture and give an upper bound for their magnetic moment. 

We initially create an ultracold two-species Fermi-Fermi mixture by sympathetic cooling of the fermionic species $\Li$ and $\K$ with an evaporatively cooled bosonic species, $\Rb$, as described previously \cite{Taglieber2006,Taglieber2008}. During the cooling process, the three species are confined in a magnetic trap in their most strongly confined and collisional stable states $\Rb\left|\mathrm{F}=2,\mathrm{m_{F}}=2\right\rangle$, $\K\left|9/2,9/2\right\rangle$, and $\Li\left|3/2,3/2\right\rangle$. For the exploitation and study of Fesh\-bach resonances (FR), the apparatus was extended by an optical dipole trap (ODT) and a setup that allows us to apply a stable homogeneous magnetic field (FB field) of up to 1\,kG in the horizontal plane. The ODT is realized by two perpendicular laser beams with the two foci coinciding at the center of the magnetic trap. The first (second) beam points along the horizontal (vertical) axis and has a $1/e^2$-radius of $55\,\mathrm{\mu m}$ ($50\,\mathrm{\mu m}$). The two beams originate from a single-mode, single-frequency ytterbium fiber laser operating at 1064\,nm. They have perpendicular polarizations and a difference frequency of $220\,\mathrm{MHz}$. The FB field is calibrated with known hyperfine and Zeeman transitions. The overall field uncertainty (including magnetic field inhomogeneities, external field fluctuations and long-term drifts) is smaller than $7\,\mathrm{mG}$. 
After sympathetic cooling of $\Li$ and $\K$ into the quantum-degenerate regime by complete evaporation of $\Rb$, any residual $\Rb$ atoms in the $\left|\mathrm{F}=2\right\rangle$ manifold are removed by a resonant light pulse. 
The Fermi-Fermi mixture is then adiabatically transferred into the ODT.

Subsequently, first $\Li$ and then $\K$ are transferred into the absolute ground states $\LiOneOne$ and $\KNinemNine$ by an adiabatic rapid passage (ARP) at $20.6\,\mathrm{G}$. This order is beneficial since, as a consequence of the inverted hyperfine structure of $\K$, it suppresses losses due to interspecies spin relaxations. Transfer efficiencies of the ARPs are almost complete for both species and the non-transferred fractions are below detection threshold. The mean trapping frequencies in the nearly isotropic ODT are $\bar{\nu}_\mathrm{Li}=1245(30)\,\mathrm{Hz}$ for $\Li$ and $\bar{\nu}_\mathrm{K}=725(20)\,\mathrm{Hz}$ for $\K$. In this trap, the typical atom numbers of $N_\mathrm{Li}\approx N_\mathrm{K} \approx 1\times 10^5$ and temperatures of $T_\mathrm{Li}=0.4\,T^\mathrm{Li}_\mathrm{F}$ and $T_\mathrm{K}=0.6\,T^\mathrm{K}_\mathrm{F}$ ($T_{\mathrm{F}}$ being the Fermi temperature) correspond to peak densities of $n_\mathrm{Li}=2.0\times 10^{13}\,\mathrm{cm}^{-3}$ and $n_\mathrm{K}=1.5\times 10^{14}\,\mathrm{cm}^{-3}$. 

In the following, we will discuss measurements with the $\Li$-$\K$ mixture in the vicinity of the interspecies FR at a magnetic field $B_0$ of about $155.1\,\mathrm{G}$ \cite{Wille2008}. 

\begin{figure}
 \centering
 \includegraphics[width=8cm]{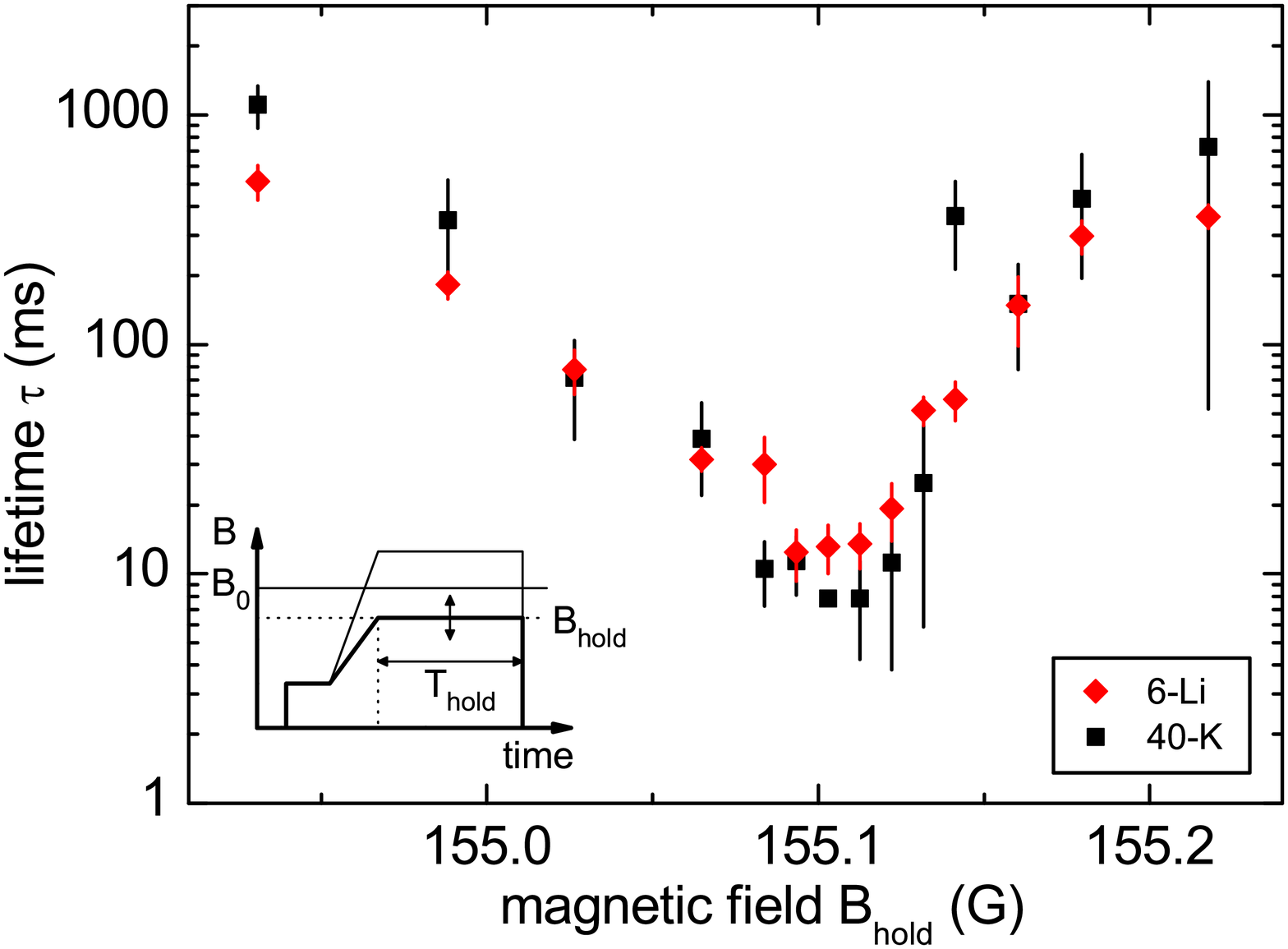}
 \caption{\label{fig:LossFeature} Lifetime of the $\Li$-$\K$ mixture as a function of the magnetic field in the vicinity of a FR between $\LiOneOne$ and $\KNinemFive$.}
\end{figure}

In a first experiment, we investigate the lifetime of the mixture close to the FR as a function of the magnetic field (see Fig.~\ref{fig:LossFeature}). Starting from the optically trapped ultracold mixture of $\Li$ and $\K$ atoms in the lowest hyperfine states, the magnetic bias field is linearly ramped within $30\,\mathrm{ms}$ from $20.6\,\mathrm{G}$ to a constant value of $152.78\,\mathrm{G}$, i.e.~to the repulsive side below the FR. The atoms are then transferred into the hyperfine states $\KNinemFive$ and $\LiOneOne$ by means of an ARP. In a second linear ramp, the magnetic field is increased within $0.5\,\mathrm{ms}$ to a variable value $B_\mathrm{hold}$, and kept constant for a duration $T_\mathrm{hold}$. It is then switched off rapidly with an initial slope of $820\,\mathrm{G/ms}$. The quantization axis is maintained by a $1\,\mathrm{G}$ bias. After a subsequent holding time of $5\,\mathrm{ms}$, the atoms are released from the ODT. Finally, the lithium and potassium clouds are detected by absorption imaging after $1\,\mathrm{ms}$ and $4\,\mathrm{ms}$ of free expansion. In order to determine the lifetime $\tau$ of the gas for a given magnetic	 field $B_\mathrm{hold}$, the experiment is repeated for at least eight different holding durations $T_\mathrm{hold}$ and an exponential decay function is fitted to the obtained atom numbers. As shown in Fig.~\ref{fig:LossFeature}, lifetime $\tau$ of the mixture as a function of $B_\mathrm{hold}$ decreases by two orders of magnitude to a minimum of $10\,\mathrm{ms}$ at $155.10(5)\,\mathrm{G}$. This loss feature observed in our experiment has a 3\,dB width of about $50\,\mathrm{mG}$. Note that this value is significantly smaller than the one obtained in Ref.~\cite{Wille2008} for the same FR but at much higher temperature. We attribute the observed decrease in atom number to the creation of molecules, which are not detected by the imaging procedure, and to losses of atoms and molecules from the trap due to vibrational relaxation. The observed weak asymmetry of the line shape is in qualitative agreement with predictions for three-body relaxation \cite{Incao2006}.

\begin{figure}
 \centering
 \includegraphics[width=8cm]{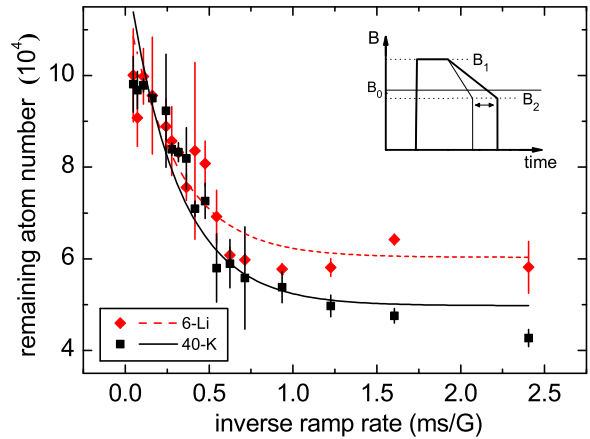}
 \caption{\label{fig:RampSpeed} Adiabaticity of the molecule conversion process as a function of the inverse sweep rate of the magnetic field strength. The decrease in atom number is attributed to molecule association. The lines represent an exponential fit based on the Landau-Zener theory.}
\end{figure}
 
In the next measurement, we convert the two-species mixture to heteronuclear molecules by sweeping the magnetic field strength across the FR. We study the adiabaticity of the association process as a function of the sweep rate $\dot{B}$ of the magnetic field strength. Starting with the two-species mixture in the lowest hyperfine states at 20.6\,G in the ODT as before, the magnetic bias field is increased within $30\,\mathrm{ms}$ to a value $B_1=156.81\,\mathrm{G}$, i.e. to the atomic side of the FR far away from the resonance. During a subsequent holding duration of $30\,\mathrm{ms}$, the atoms are transferred into the states $\KNinemFive$ and $\LiOneOne$. The magnetic field strength is then linearly ramped with a variable ramp speed to a fixed final value $B_2=154.73\,\mathrm{G}$ far on the molecular side of the FR. As before, the magnetic bias field is then rapidly switched off and the remaining atoms are detected after 5\,ms holding time in the ODT and subsequent free expansion. Since the imaging procedure at low magnetic field detects only free atoms, the creation of molecules shows up in Fig.~\ref{fig:RampSpeed} as a decrease in the observed atom number. In this way we typically detect $4 \times 10^4$ molecules with a conversion efficiency close to $50\,\%$. By fitting an exponential function to the atom numbers, we obtain a characteristic inverse sweep rate of $1/\dot{B}_\mathrm{ad}=0.30(5)\,\mathrm{ms}/\mathrm{G}$ for adiabatic association of atoms into molecules.

\begin{figure}
 \centering
 \includegraphics[width=8cm]{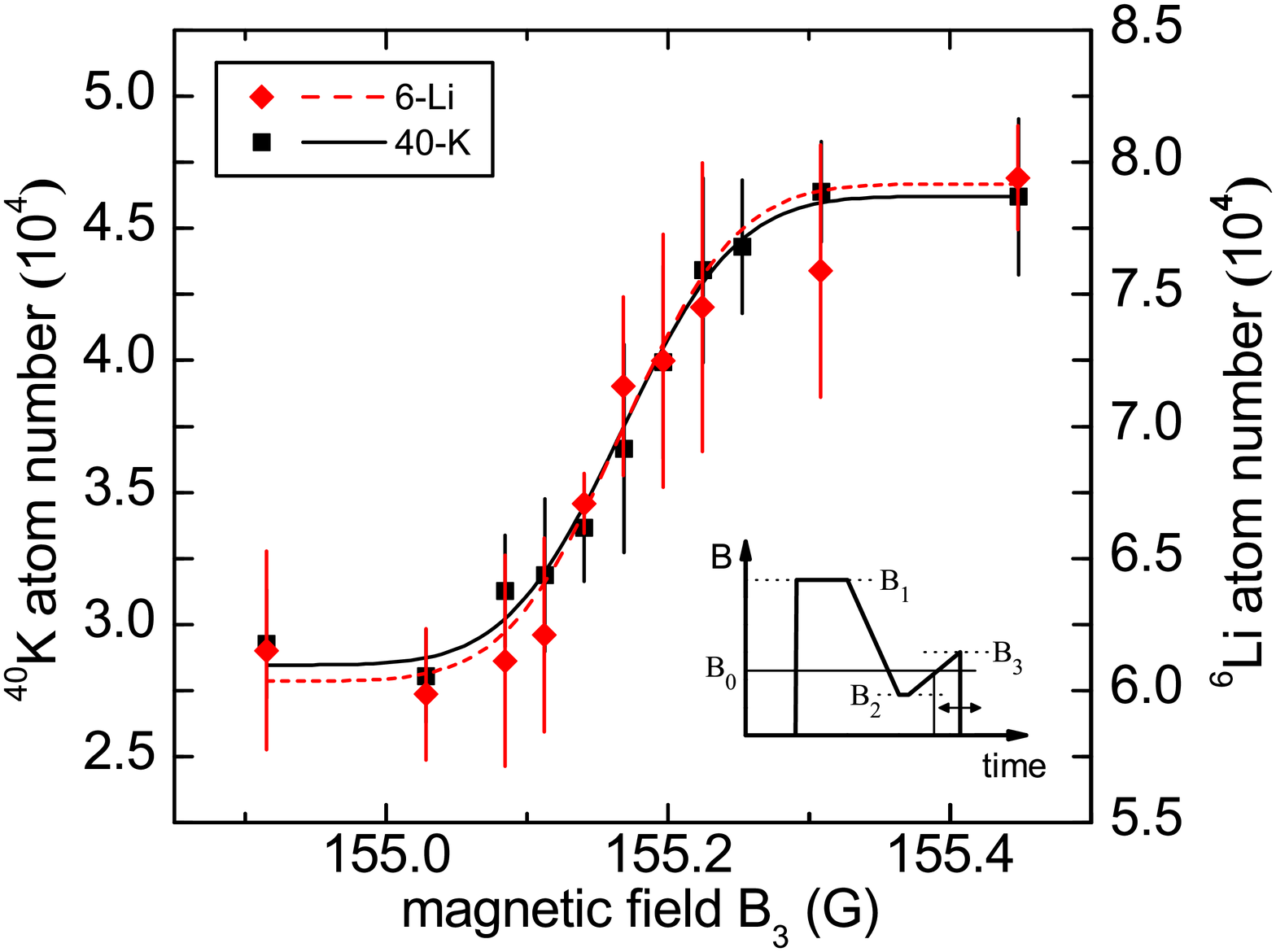}
 \caption{\label{fig:SweepDownAndUp} Reconversion process of molecules to unbound atoms as a function of the final magnetic field strength. First, molecules are associated by an adiabatic sweep from $B_1$ to $B_2$. Then, the molecules are adiabatically reconverted to unbound atoms by a sweep to a variable final value $B_3$. The lines represent an empirical fit using an error function.}
\end{figure}

Next, we show that the observed decrease in atom number can indeed be attributed to the creation of molecules (see Fig.~\ref{fig:SweepDownAndUp}). For this purpose, the atoms are prepared in the ODT in the states $\KNinemFive$ and $\LiOneOne$ at the magnetic field strength $B_1$ as in the previous measurement. The magnetic field strength is then linearly ramped with a constant rate of $1\,\mathrm{G}/\mathrm{ms}$ to the value $B_2$ on the molecular side of the FR, thus converting free atoms into weakly bound molecules. After $200\,\mu\mathrm{s}$, the magnetic field strength is linearly ramped back with the same sweep rate to a variable final value $B_3$. As before, the magnetic field is then rapidly switched off and the atoms are detected after 5\,ms holding time in the ODT and subsequent free expansion. Fig.~\ref{fig:SweepDownAndUp} shows the lithium and potassium atom numbers as a function of the final magnetic field $B_3$. Towards higher final magnetic fields $B_3$, molecules are dissociated back into atoms resulting in atom count for both species that is increased by $1.8(3)\times 10^{4}$. Note that increase does not reflect the total number of molecules created, as inelastic loss of the molecules can occur before the dissociation sweep. By an empirical fit to the number of detected atoms, we find the reconversion process to be centered at $155.17(8)\,\mathrm{G}$, where the uncertainty represents the 10\% and 90\% levels. We also created molecules at another, narrower FR between the states $\KNinemNine$ and $\LiOneOne$, which we locate at $168.6(2)\,\mathrm{G}$. For this resonance, we obtain up to $2.0(5) \times 10^4$ molecules.

For the following series of measurements, we work at lower densities in a weaker ODT with mean trapping frequencies of $\bar{\nu}_\mathrm{Li}=347(8)\,\mathrm{Hz}$ for lithium and $\bar{\nu}_\mathrm{K}=203(5)\,\mathrm{Hz}$ for potassium. 

\begin{figure}
 \centering
 \includegraphics[width=8cm]{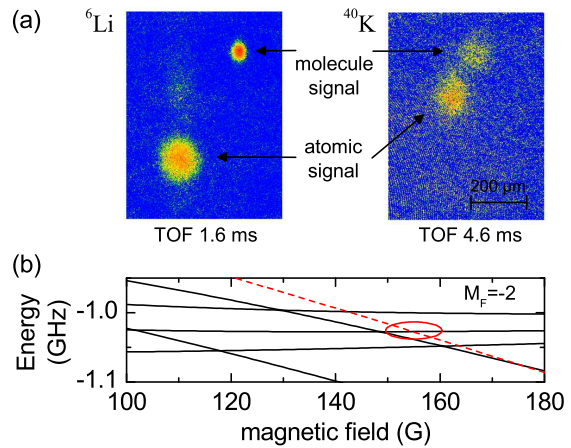}
 \caption{\label{fig:SG} (a) Direct simultaneous absorption imaging of the molecular and atomic cloud after separation by a Stern-Gerlach force during time of flight (TOF). (b) Energies of molecular states (solid lines) and free atoms (dashed line): The molecular state has almost vanishing magnetic moment at the Feshbach resonance (encircled region).}
\end{figure}

Direct evidence of the molecules and purification can be obtained by spatially separating  the atoms from the molecular cloud by applying a Stern-Gerlach force before imaging. For this purpose a difference of the magnetic moments of the molecules, $\mu\!\left(\Li\K\right)$, and the free atoms, $\mu\!\left(\Li\right)$ and $\mu\!\left(\K\right)$, is exploited. This difference can be inferred from the magnetic field dependence of the molecular bound state energy as compared to the energies of the two free atoms, as shown in Fig.~\ref{fig:SG}(b). The dependence of the molecular energy is calculated with the asymptotic bound state model \cite{Wille2008} for the manifold with a total projection quantum number $M_{F}=-2$, which is relevant in our case.
For this measurement, we first prepare the molecules at $155.03\,\mathrm{G}$. Then the ODT is switched off and a magnetic field gradient pulse with an average strength of $167\,\mathrm{G/cm}$ is applied for $570\,\mathrm{\mu s}$. After a total time of free expansion of $1.6\,\mathrm{ms}$ for $\Li$ and $4.6\,\mathrm{ms}$ for $\K$, the atomic and molecular clouds are simultaneously imaged using a closed optical transition at the constant high magnetic FB field. Simultaneous imaging of molecules and atoms near the FR is facilitated in the heteronuclear case by the fact that the ground and the first excited states share the same asymptotic behavior \cite{Ospelkaus08}. The resulting images presented in Fig.~\ref{fig:SG}(a) directly show the molecular cloud clearly separated from the atomic cloud. The different masses of the molecules and atoms result in different expansion velocities of the corresponding clouds, even at equal temperature. From measuring the trajectory of the molecules in a variable magnetic field gradient we determine a negligible magnetic moment of the molecules of $<\!0.1\,\mathrm{\mu_{B}}$, as expected from the asymptotic bound state model.

\begin{figure}
 \centering
 \includegraphics[width=8cm]{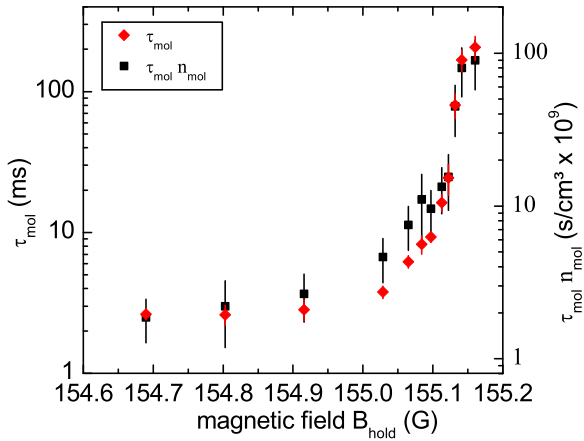}
 \caption{\label{fig:Lifetimemol} Characterization of the molecular $1/e$ lifetime $\tau_{\mathrm{mol}}$ in the molecule-atom mixture as a function of the magnetic field strength. In addition, $\tau_{\mathrm{mol}}\,n_{\mathrm{mol}}$ is plotted, where $n_{\mathrm{mol}}$ is the initial molecular density, averaged over the cloud.}
\end{figure}

Furthermore, we measure the $1/e$ lifetime of the molecules $\tau_{\mathrm{mol}}$ in the molecule-atom mixture as a function of the magnetic field strength (see Fig.~\ref{fig:Lifetimemol}). The heteronuclear molecules are created by an adiabatic linear ramp of the magnetic field strength from $156.81\,\mathrm{G}$ to a variable final value $B_\mathrm{hold}$. For each value of $B_\mathrm{hold}$, the molecule-atom mixture is held for a variable duration in the weak ODT before the molecular and atomic clouds are released from the trap and detected separately as described in the previous measurement. For this measurement, the peak densities before the magnetic field sweep are $n_\mathrm{Li}=2.9\times 10^{12}\,\mathrm{cm}^{-3}$ and $n_\mathrm{K}=2.2\times 10^{13}\,\mathrm{cm}^{-3}$ with temperatures $T_{Li}=0.3\, T^\mathrm{Li}_\mathrm{F}$ and $T_{K}=0.4\, T^\mathrm{K}_\mathrm{F}$. We observe lifetimes $\tau_{\mathrm{mol}}$ of the heteronuclear molecules of more than $100\,\mathrm{ms}$. Importantly, we find that $\tau_{\mathrm{mol}}$ is increased by almost two orders of magnitude for magnetic field strengths close to the resonance compared to the value far away from the resonance on the molecular side. This observation is consistent with previous results for homonuclear fermionic spin mixtures \cite{HomonuclearFermiMolecules, StabilityFermiFermiMixture}. Since due to the molecule creation process the densities of the molecules and atoms depend on $B_\mathrm{hold}$, we have to show that the increased lifetime close to the resonance is not only a mere consequence of a smaller molecule density. For this purpose, we also plot the product of the lifetime and the initial average density of the molecules, $\tau_{\mathrm{mol}}\,n_{\mathrm{mol}}$, in Fig.~\ref{fig:Lifetimemol}. We find that also this parameter is significantly increased close to resonance. Note that such suppression of inelastic decay has been explained on the basis of the Pauli principle for the case of an open-channel dominated FR between fermionic atoms \cite{Petrov2005}. However, the FRs used in this work are expected to be closed-channel dominated \cite{Wille2008}.

The initial temperatures for $\Li$ and $\K$ are smaller than the predicted critical temperature $T_{\mathrm{c}}=0.52\,T_{\mathrm{F}}$ for conversion into a heteronuclear molecular BEC. While this has not been observed, the long molecular lifetime is a good basis to enter this regime by additional cooling. Further, a simple estimation based on our experimental parameters and on the width of the FR given in Ref.~\cite{Wille2008} shows that we can prepare the Fermi-Fermi two-species mixture completely in the strongly interacting regime ($k_\mathrm{F}\left|a\right|\geq 1$, where $k_\mathrm{F}$ is the Fermi wave vector and $a$ the s-wave scattering length). This, in combination with the long lifetime, may allow us to study many-body physics of a heteronuclear mixture at a closed-channel FR.

In conclusion, we have created the first heteronuclear bosonic molecules consisting of two fermionic atomic species. We used direct imaging of the molecules as a sensitive probe to map the lifetime of the molecule-atom mixture. We found increased lifetimes of the molecules of more than $100\,\mathrm{ms}$ close to resonance as expected for fermions on grounds of the Pauli principle. Therefore, this is the first system that can be used to explore many-body physics of a heteronuclear mixture with two different masses in the strongly interacting regime across a Feshbach resonance. The molecular cloud being close to quantum degeneracy marks the first starting point for the realization of a heteronuclear molecular BEC and for transfer into the vibrational ground state \cite{Ni2008} to create a polar BEC. Another intriguing goal could be the exploration of asymmetric Cooper pairs.

This work was supported by the German Science Foundation (DFG) via Research Unit FOR801 and the Munich Center for Advanced Photonics.

{\em Note added:} Recently, we became aware of an experiment with heteronuclear bosonic molecules from a Bose-Bose mixture (${}^{41}\textrm{K}$-${}^{87}\textrm{Rb}$) \cite{Weber2008}.


\end{document}